\documentclass{article}

\usepackage{arxiv}
\usepackage[utf8]{inputenc}	
\usepackage[T1]{fontenc}		
\usepackage{hyperref}		
\usepackage{url}				
\usepackage{booktabs}		
\usepackage{amsfonts}		
\usepackage{nicefrac}		
\usepackage{microtype}		
\usepackage{graphicx}
\usepackage{tabularx}

\title{Drawing inspiration from biological dendrites to empower artificial neural networks}

\date{}

\author{ Spyridon Chavlis \\
	Institute of Molecular Biology and Biotechnology (IMBB)\\
	Foundation for Research and Technology Hellas (FORTH)\\
	Plastira Avenue 100, Heraklion, 70013, Greece \\
	\texttt{schavlis@imbb.forth.gr} \\
	\And
Panayiota Poirazi \\
	Institute of Molecular Biology and Biotechnology (IMBB)\\
	Foundation for Research and Technology Hellas (FORTH)\\
	Plastira Avenue 100, Heraklion, 70013, Greece \\
	\texttt{poirazi@imbb.forth.gr} \\
}

\renewcommand{\headeright}{Opinion article}
\renewcommand{\undertitle}{Opinion article}
\renewcommand{\shorttitle}{Biological dendrites in artificial neural networks}

\hypersetup{
pdftitle={Drawing inspiration from biological dendrites to empower artificial neural networks},
pdfsubject={q-bio.NC},
pdfauthor={Spyridon Chavlis, Panayiota Poirazi},
pdfkeywords={Artificial Neural Networks, Biological dendrites, Plasticity},
}

\begin{document}
\maketitle

\begin{abstract}
This article highlights specific features of biological neurons and their dendritic trees, whose adoption may help advance artificial neural networks used in various machine learning applications. Advancements could take the form of increased computational capabilities and/or reduced power consumption. Proposed features include dendritic anatomy, dendritic nonlinearities, and compartmentalized plasticity rules, all of which shape learning and information processing in biological networks. We discuss the computational benefits provided by these features in biological neurons and suggest ways to adopt them in artificial neurons in order to exploit the respective benefits in machine learning.
\end{abstract}


\section{Introduction}
Artificial Neural Networks (ANN) implemented in Deep Learning (DL) architectures have been extremely successful in solving challenging machine learning (ML) problems such as image \cite{LeCun2015}, speech \cite{Nassif2019}, and face \cite{Masi2018} recognition, playing online games \cite{Justesen2020}, autonomous driving \cite{Huang2020}, etc. However, despite their huge successes, state-of-the-art DL architectures suffer from problems that seem rudimentary to real brains. For example, while becoming true experts in solving a specific task, they typically fail to transfer these skills to new problems without extensive retraining – a property known as “transfer learning”. Moreover, when learning new problems, they tend to forget previously learned ones -a problem termed “catastrophic forgetting.” Finally, in order to achieve top performance, they typically require large training sets and thousands/millions of trainable parameters, necessitating the use of computing systems that leave a substantial carbon footprint. The brain, on the other hand, runs on extremely low energy (< 20 watts), learns from few examples, generalizes extremely well, and can learn in a continual manner without erasing useful prior knowledge. In this article, we argue that increasing the bioinspiration of ANNs, by adopting dendritic features may help address the abovementioned limitations. We discuss two types of benefits: advancing the computational power of ANNs and/or reducing their power consumption. \\

\textbf{\textit{Why dendrites?}}

\begin{figure}
	\centering
	\centerline{\includegraphics[scale=0.60]{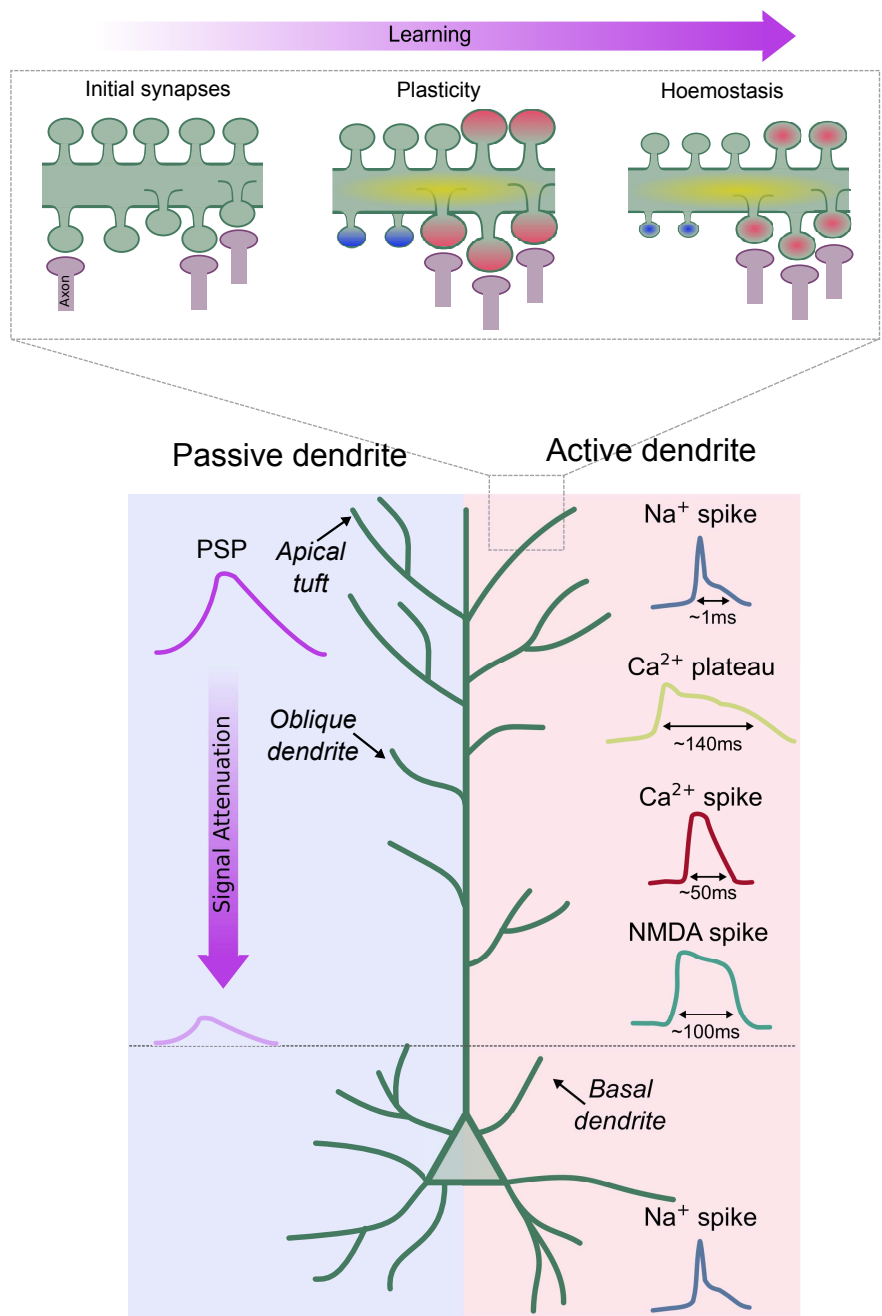}}
	\caption{Schematic representation of dendritic features in biological neurons. (Bottom) Schematic depiction of a typical pyramidal cell. If dendrites were passive (blue shaded area, left), the magnitude of an incoming signal would decrease as it propagates to the soma. However, with active dendrites (red-shaded area, right), a strong-enough incoming stimulus can generate various types of dendritic spikes (‘dSpikes’). Typical shapes of ‘dSpikes’ are shown schematically on the right (blue: Na\textsuperscript{2+} spike (duration: 1-2ms), red: Ca\textsuperscript{2+} spike (duration: a few 10s of milliseconds), green: NMDA spike (duration: 50-150ms), yellow: Ca\textsuperscript{2+} plateau potentials, which are mixtures of Ca\textsuperscript{2+} and NMDA spikes). (Top) Schematic representation of some plasticity mechanisms that operate in dendrites. (left) Initial connectivity, all synapses are of the same strength. (middle) Some of the synapses (the co-active ones) are strengthened, while others are depressed. The overall dendritic excitability increases (yellow shade). Axons that contact depressed spines rewire, and new connections are formed. (right) Homeostasis scales all synapses while maintaining their relative strengths.}
	\label{fig:fig1}
\end{figure}

Dendrites are thin processes that extend from the cell bodies of neurons and serve as their primary receiving end: over 90\% of the synapses in principal neurons are localized within their dendrites. Dendrites come in a great variety of shapes depending on the species, brain area, cell type, and even the layer in which they reside, indicating that they may be specialized for the respective computations needed \cite{Stuart2016}. Beyond their function as entry devices, morphologically complex dendrites affect the integration and propagation of incoming signals in complex ways. For example, passive dendrites act as filters due to the attenuation and temporal filtering of signals traveling to the cell body \cite{Spruston1994}, thus allowing neurons to infer the distance of incoming signals from the soma and their kinetics. Moreover, active dendrites are home to a variety of voltage-gated conductances that support the generation of dendritic spikes (reviewed in \cite{Major2013}, Figure \ref{fig:fig1}). These dendritic features amount to different types of nonlinear integration and allow individual neurons to perform complex computations, ranging from logical operations to signal amplification or segregation, temporal sequence/coincidence detection, parallel nonlinear processing, etc. \cite{Ariav2003, Branco2011, Polsky2004, Softky1994,Spruston2008,Ujfalussy2009, Ujfalussy2018}.

The computational capabilities of neurons with nonlinear dendrites are exploited by a set of multiscale plasticity rules that induce synaptic, intrinsic, homeostatic, and wiring changes \cite{Bono2017,Kastellakis2019}. These rules can differ between neuronal (e.g., apical vs. basal trees) and/or dendritic (e.g., proximal vs. distal) compartments \cite{Gordon2006, Magee2020,Mel2017,Sajikumar2011,Weber2016}, enabling neurons to alter dendritic integration in a targeted, non-uniform manner. For example, dendrites whose spikes do not propagate effectively to affect somatic output can turn into powerful influencers as a result of branch-strength-potentiation, a form of local intrinsic plasticity \cite{Losonczy2008}.

Importantly, the effects of anatomical, biophysical, and plasticity properties of dendrites in the living animal are often tangled. For example, considering only the anatomy of the dendritic tree, distant inputs should be severely attenuated before reaching the soma, causing weak somatic depolarization. However, this is not always the case. Voltage- and ligand-gated channels in the distal dendrites are activated upon the arrival of synchronous inputs, leading to the generation of dendritic spikes that reliably propagate to the soma, often causing somatic spikes (Figure \ref{fig:fig1}). Because of these nonlinear interactions, biological neurons are considered much more powerful than linear ANN nodes (or point neurons) \cite{Jadi2014,London2005,Poirazi2003a,Tzilivaki2019,Ujfalussy2018}. Instead, dendrites rather than neurons are proposed to serve as the fundamental functional unit of the brain \cite{Branco2010}. Of note, the nonlinear computations performed by neurons with active dendrites can also be performed by networks of point neurons. However, solving them within individual neurons provides important savings: more energy would be required to fire APs across many neurons and to form appropriate axonal wiring/coordination of these neurons. Moreover, the space taken up by neuronal somata is much larger than the space occupied by dendrites, therefore computing with dendrites is more resource-efficient for biological brains.

The potential benefits of the abovementioned dendritic characteristics have yet to be exploited in ANNs and their neuromorphic implementations. Respective neurons (nodes) in traditional ANNs are typically point neurons arranged in several layers, interconnected with symmetric (all-to-all), plastic (trainable) synapses, and furnished with nonlinear activation functions. Learning is typically achieved through weight updating rules, the most popular of which is the backpropagation-of-error algorithm. There are strengths in the simplicity of this approach (easier to understand and implement, fast convergence, high performance), but also weaknesses (e.g., catastrophic forgetting, high power consumption). To address the latter, researchers turn to more biologically-inspired architectures \cite{Guerguiev2017,Kreiman2020,Payeur2021,Siegel2000} and learning rules \cite{Kirkpatrick2017,Kording2001,Payeur2021}, with promising results (e.g., on MNIST classification). We believe that an in-depth exploration of the anatomical, biophysical, and plasticity features of dendrites may help reduce the resources needed to achieve top performance and/or address problems like transfer and continual learning, where traditional ANNs fail.
Towards this goal, we review recent experimental and computational findings related to how dendritic features can form a foundation for specific computations and suggest ways to integrate them in DL architectures. We focus on three fundamental features of dendrites: i) anatomy and compartmentalization, ii) active ionic mechanisms, and iii) multiscale plasticity rules. We spotlight recent discoveries that connect subcellular mechanisms and algorithmic-level functions for each one of these dendritic features.

\section{Dendritic anatomy and compartmentalization}
\label{sec:anatomy}

The pioneering work of Wilfrid Rall (1959) revealed that dendrites are not isopotential electrical devices but rather electrically distributed circuits whose conducting properties are described by the cable theory. As such, signals that travel along (passive) dendrites are logarithmically attenuated \cite{Zador1995}, and the level of this attenuation is highly dependent on the morphological properties of dendritic trees \cite{Spruston1994,Stuart1998}. Specifically, the propagation loss is proposed to decrease with increasing diameter and decreasing length of dendritic compartments \cite{Baruah2019}. Given that dendritic trees are typically non-symmetric, signal attenuation can vary extensively within the same neuron. Moreover, branch points are a major source of attenuation, a feature that promotes compartmentalized processing in neurons with elaborated trees \cite{Ferrante2013}. These morphological effects have been attributed important functions: according to Rall’s 1964 study \cite{Rall1964}, passive dendrites can detect the temporal sequence of incoming inputs by inducing different somatic responses depending on the activation order. This property could be used by neurons to infer the direction of motion (e.g., of a visual stimulus moving from right to left versus from left to right). Moreover, the compartmentalized, sublinear integration of inputs in passive dendrites was suggested to enable single neurons to compute linearly non-separable functions \cite{Caze2013}. Finally, the dendritic structure has been shown to enhance sparsity in biological networks, a property associated with increased memory capacity \cite{Brunel2004}, better discrimination \cite{Chavlis2017}, and faster learning \cite{Schweighofer2001}. For example, dendrites made granule cells less likely to fire in a dentate gyrus network model. Consequently, the network’s ability to discriminate between similar input patterns was increased because patterns were encoded by less overlapping granule cell populations \cite{Chavlis2017}. Moreover, sparsifying the connectivity of inputs, by distributing them across basal and apical dendrites, enabled simulated neurons to encode multiple input sequences simultaneously \cite{Hawkins2016}. These examples highlight the power of dendritic anatomy in shaping signal processing in biological neurons, primarily by implementing sublinear integration and increasing network sparsity.

ANNs, on the other hand, typically use the McCulloch-Pitts model \cite{McCulloch1943}, whereby artificial neurons (nodes) receive a weighted sum of their incoming synaptic inputs, largely ignoring dendritic influences. Yet, some of the approaches commonly used in the field to improve performance resemble the structural features of dendritic trees. For example, dropout layers used in deep architectures \cite{Srivastava2014} lead to the sparsification of a network, as a randomly chosen portion of the connections are removed from training. Such sparsification is somewhat similar to that achieved by having multi-branched dendritic trees and achieves both resource savings (fewer trainable parameters) and better generalization performance. Modular networks \cite{Poirazi2004, Shukla2010}, consisting of various ANN sub-networks, can also be seen as analogous to dendritic structures, with each module learning a part of the task. Finally, sparse ANNs, where a portion of the trainable parameters are set to zero (and respective multiplications can be omitted), have some similarities with the restricted connectivity imposed by dendritic structures \cite{BaoyuanLiu2015}.

\section{Active ionic mechanisms and dendritic computations}
\label{sec:active}

In addition to their branching structure, many dendrites are equipped with a variety of voltage-gated and ligand-gated ionic channels. These so-called “active” mechanisms allow dendrites to transform incoming signals in nonlinear ways. For example, if two stimuli arrive within the same dendrite of pyramidal neurons, their integration results in a much larger response than if they land in separate dendrites \cite{Losonczy2006,Poirazi2003,Polsky2004}. This is because the integration of multiple nearby inputs causes a large enough depolarization that activates voltage-gated channels, which would otherwise remain closed, resulting in the generation of regenerative events (dendritic spikes). Dendritic spikes are much larger in amplitude than the linear summation of the incoming synaptic inputs (thus resulting in signal amplification) and can reliably propagate to the cell body, counteracting the attenuation caused by dendritic morphology. The ability to induce different responses depending on the dendritic location of inputs greatly increases the discriminatory power of neurons \cite{Poirazi2001,Poirazi2003a}. 
Dendritic spikes are driven by various ions and differ in their kinetics and localization. Dendritic sodium (Na\textsuperscript{+}) spikes are faster than calcium (Ca\textsuperscript{2+}) spikes and occur throughout the dendritic tree \cite{Ariav2003,Golding1998,Milojkovic2005,Spruston2008}. Ca\textsuperscript{2+} spikes are generated in the distal apical trunk and can spread to the apical dendritic tree \cite{Larkum2009,Schiller1997}. NMDA spikes have even slower dynamics and usually reside within side/terminal dendritic branches \cite{Cichon2015}. Finally, Na\textsuperscript{+} spikes generated at the soma can also travel back into the dendrites \cite{Brunner2016}. The occurrence of these dendritic spikes, their kinetics, and location dependence expand the computing power of individual neurons \cite{Silver2010}. For example, both Na\textsuperscript{+} and Ca\textsuperscript{2+} spikes have been attributed to a coincidence detector role. The former were shown to emerge upon synchronous activation of synaptic inputs within an individual dendrite \cite{Ariav2003,Losonczy2006}, thus enabling a neuron to detect synchronous inputs with millisecond precision. The latter were shown to emerge when distal and proximal dendritic regions of cortical neurons are activated within several milliseconds, leading to strong somatic bursting \cite{Larkum1999}. Bursting can signal the coupling (coincidence) of input streams from different regions, thus serving as an association mechanism operating at larger time scales. Finally, NMDA spike generation within individual branches was found to be sensitive to both the timing and location of inputs, thus endowing single dendrites with the ability to detect the activation sequence of inputs \cite{Branco2011}.

Beyond their spatio-temporal processing power, dendritic spikes were predicted by models \cite{Poirazi2003a, Poirazi2003} and verified by experiments \cite{Losonczy2006,Polsky2004} to enable dendrites to integrate inputs like sigmoidal activation functions. As a result, a pyramidal neuron with multiple dendrites (sigmoidal units) acts like a powerful 2-stage \cite{Jadi2014, Poirazi2003a} or even a multi-stage ANN \cite{Beniaguev2020, Hausser2003}, with dendrites acting as computing nodes. Such dendritic nonlinearities were recently found in interneurons \cite{Chiovini2014,Katona2011,Tran-Van-Minh2016}, suggesting that these cells may also act as multi-stage ANNs \cite{Tzilivaki2019}. Notably, Gidon et al. \cite{Gidon2020} recently discovered a new, more powerful dendritic spike (dCaAP) in human layer 2/3 cortical neurons. Unlike other types of spikes, the input-output curve for dCaAPs is described by a non-monotonic activation function. In biophysical models, this property enables a single human dendrite to solve the Exclusive-OR (XOR) problem, a logical operation though solvable only by multilayer ANNs. In sum, dendritic spikes endow dendrites with nonlinear signal processing capabilities, which enable the local implementation of various nonlinear computations and can be mathematically approximated with activation functions similar to those used in ANNs. Of note, the suppressing effects of passive dendrites (sublinear integration) can also be described with activation functions.

In the ML world, activation functions are a topic of intense investigation because of their enormous impact on the performance accuracy of ANNs \cite{Hayou2019}. Currently, the most popular activation function is the ReLU and its derivatives (e.g., PReLU, ELU, GELU), although more biological ones (e.g., sigmoid, tanh, delta, logarithmic) are also extensively used. The positive part of ReLU can be thought of as the frequency-current (F/I) curve of a neuron, while the negative part (e.g., leaky ReLU) is there for computational efficiency.

A number of recent ML studies have explored dendritic structure and activation functions in ANNs. For example, \cite{Guerguiev2017} introduced three-compartment-nodes comprised of a soma, an apical, and a basal dendrite, each furnished with its own activation function. Basal and apical compartments received different types of information (sensory and feedback, respectively), allowing neurons in different layers to coordinate synaptic weight updates. As a result, the network learned to categorize images better than a single-layer network. While this work aimed at understanding how the compartmentalization of biological neurons may facilitate learning, it is one of the first studies where simplified dendrites were integrated as building blocks of the ANN graph. A similar approach using dendritic ANNs, i.e., replacing each node with a 2-layered structure furnished with a dendritic and somatic layer, achieves higher model expressivity (complexity of functions that can be computed) compared to that of a single-layered neuron and better generalization than traditional ANNs, by sparsifying the network connections \cite{Wu2018}. Jones and Kording \cite{Jones2020} went a step further by incorporating multiple dendritic layers in an ANN. They showed that a single neuron modeled as a dendritic ANN with sparse connectivity could solve binary classification tasks (e.g., MNIST) and benefit from having the same input at different dendritic sites. Another approach utilizes multiple dendritic layers in each node of a deep ANN, and reduces catastrophic forgetting in a continual learning task \cite{Camp2020}. To explore the potential advantages of human dCaAPs, a recent study used their respective transfer function in various deep learning architectures and found a small but significant increase in classification performance for various tasks, especially when used in combination with sparse connectivity resembling that of dendritic trees \cite{Georgescu2020}. It should be noted that none of the abovementioned studies has surpassed the current state-of-the-art in performance accuracy. However, it has been shown that dendritic ANNs can achieve better performance (lower error) when compared with traditional ANNs with the same number of trainable parameters \cite{Wu2018}.

While the true power of dendrites for ML is just beginning to emerge, we believe that their most prominent role is to allow neurons to solve difficult problems more efficiently. By efficiency, we refer to the computational cost of solving a problem. We can simplify this to the number of parameters that must be updated and the energy that must be consumed to make all necessary computations and updates. For example, a human neuron can solve XOR using solely its apical dendrites, whereby a non-monotonic function resides \cite{Gidon2020}. With point-neurons, this would require a small network of linear-non-linear units. Obviously, the biological solution is cheaper, therefore, more efficient (no need to engage multiple neurons, which have a large volume, and it is hard to activate them nor update their connection weights). Another example comes from associative memory formation: a memory stored in a network of dendritic neurons is encoded by fewer neurons that fire at lower rates, compared to a memory stored in a network of point neurons. At the same time, dendrites maximize the sparsity of responses and minimize memory interference \cite{Poirazi2001, Tzilivaki2019}. This is achieved using the exact same number of trainable parameters (synaptic weights) for both types of networks. Overall, if the aforementioned benefits in resource savings could be realized in ML, they would bring major advantages in systems where resource consumption is critical (e.g., in portable devices, mobile phones, microchips, etc.). Achieving this, however, is not trivial as a one-to-one correspondence between the biological and the ANN components responsible for the observed benefits does not always exist.

\section{Multiscale plasticity rules}
\label{sec:plasticity}

The abovementioned savings in biological networks are not only a result of their dendritic structure and active mechanisms. These features act in concert with a set of plasticity rules that dynamically shape not only synaptic weights but also network connectivity and activation functions, and whose contribution to learning is often unintuitive \cite{Magee2020}. For example, synaptic plasticity can occur in the absence of somatic spiking \cite{Bittner2017,Kampa2007}, and synapses can be potentiated or depressed according to the activity of their neighbors \cite{El-Boustani2018}. As a result, synapses that fire together are more likely to cluster within the same dendrite, irrespectively of whether they fire strongly enough to excite the cell \cite{Kastellakis2016,Lee2016,Weber2016}. The latter is contrary to the common belief that neuronal firing is necessary for synaptic plasticity (as in classical -but not local- Hebbian rules) and highlights the richness of biological ways to update synaptic weights.

Importantly, the probability of finding co-active synaptic partners is ensured by rewiring, a structural plasticity mechanism that allows synapses to form and retract within dendrites in search of an appropriate partner \cite{Chklovskii2004,DeBello2008,DeBello2017,Legenstein2011,Limbacher2020}. This type of plasticity is an example of how nature implements dynamics changes in the wiring (connectivity/graph) of biological networks while at the same time facilitating synaptic weight updates. Finally, synapses also undergo homeostatic plasticity (scaling) to prevent run-away excitation (or inhibition) and maintain a high signal-to-noise ratio between potentiated synapses while allowing further plastic changes to take place \cite{Turrigiano2008}. This homeostatic plasticity can also be dendrite-specific \cite{Rabinowitch2006}. 
Beyond synaptic weight and wiring plasticity, the dendrites and somata of biological neurons also undergo plasticity of their intrinsic excitability \cite{Disterhoft2006,Losonczy2008,Ohtsuki2012}, which is analogous to changes in their activation functions. For example, learning-induced reductions in A-type K\textsuperscript{+} currents can make dendritic activation functions more supralinear and affect the coupling between dendritic and somatic compartments \cite{Losonczy2008,Makara2009}. Similarly, changes in the somatic afterhyperpolarization current of neurons involved in learning allow them to become more excitable and remain at a heightened excitability state for hours \cite{Disterhoft2006}. 
The specific roles sub-served by each plasticity type (weight, wiring, activation function) remain unclear, yet some contributions are directly relevant to ML in the sense that they could provide inductive biases. For example, the plasticity of somatic excitability increases the probability that a neuron involved in learning one piece of information will also participate in learning subsequent events (within a window of hours). This property facilitates the linking of information across time, forming memory episodes that extend over several hours \cite{Cai2016,Kastellakis2016}. Such a learning rule could, for example, empower ANNs that learn sequences (e.g., recurrent ANNs). Moreover, at the single neuron level, temporal binding of information is realized via the co-strengthening (clustering) of synapses from subsequent events in the same dendrites \cite{Kastellakis2016}. Synapse clustering results from the combined effects of rewiring and cooperative plasticity \cite{Kastellakis2016,Poirazi2001}, it can be established in development and persist through adulthood under a common framework \cite{Kirchner2019}, and is a very effective way to bind together pieces of information that arrive simultaneously from different input streams, thus forming associations \cite{Ariav2003,Losonczy2006}. Such plasticity rules facilitate the detection of regularities in the input distributions sensed by biological networks and bias the system towards learning higher-order correlations. We propose that incorporating rewiring and cooperative plasticity in dendritic ANNs is likely to facilitate learning in dynamically changing environments (such as transfer and continual learning). Recent attempts to implement rewiring in deep ANNs, based on a probabilistic approach, are highly promising with respect to performance accuracy \cite{Bellec2018}.

Finally, another important feature of biological plasticity rules is that they don’t need a signal dictating how close the network is to solve a specific problem. Associative learning, for example, can be achieved in a fully unsupervised manner \cite{Kastellakis2016}. On the other hand, cortical circuits were recently reported to implement supervised learning via the use of feedback signals \cite{Doron2020}. The ability of biological networks to implement both supervised and unsupervised learning is another property that could help empower ANNs. The latter are typically trained using the backpropagation-of-error algorithm (back-prop) although more sophisticated and complex (often bio-inspired) rules have also been applied \cite{Illing2019,Perez-Sanchez2018}. The back-prop moves along the direction of the minimal gradient, thus rapidly converging to the optimal solution. However, back-prop requires that every synapse in the multiple layers of an ANN knows its contribution to the global error, which is computed at the very last layer. Such a condition may appear biologically implausible, yet various bio-realistic solutions have been suggested \cite{Lillicrap2016,Lillicrap2020,Richards2019,Sacramento2018}. For example, according to Sacramento et al. \cite{Sacramento2018}, prediction errors are encoded in the distal dendrites of excitatory neurons through the activity of local interneurons and are mediated by local plasticity rules. In support of this idea, Doron and colleagues \cite{Doron2020} recently found that the distal dendrites of cortical pyramidal neurons receive such feedback signals that coordinate learning in ways similar to the backpropagation algorithm. Another approach uses error signals that are calculated locally, in a layer-specific manner \cite{Mostafa2018,Nokland2019}, thus eliminating the necessity to propagate the global error back to all hidden layers. The difference target propagation is another alternative \cite{Lee2015}, whereby targets instead of gradients are computed based on an autoencoding feedback loop, and thus, minimization is achieved with auxiliary variables \cite{Choromanska2019}. In other words, the problem is broken down to easier-to-solve local sub-problems via inserting auxiliary variables corresponding to activations in each layer.

Beyond the attempt to find bio-realistic solutions to error-backpropagation, other biological learning rules have recently been used in combination with dendritic structure. Combining a neuro-inspired plasticity rule based on bursting with a dendritic-like sparse connectivity structure was recently shown to achieve very good classification performance \cite{Payeur2021}. Moreover, in ANNs with multi-compartmental dendrites, an online optimization algorithm was shown to perform canonical correlation analysis, an unsupervised task where the data are projected in a new subspace maximizing their correlations \cite{Lipshutz2020}. It should be noted that incorporating more biologically plausible rules in dendritic ANNs does not necessarily improve performance per se, but it often achieves similar performance using fewer resources. For example, according to Illing et al. \cite{Illing2020}, a local, biologically plausible learning rule based on a layer-specific loss function can learn representations of images and speech with less hardware consumption. Overall, these examples pinpoint the potential benefits of incorporating bio-inspired plasticity rules, in conjunction with dendritic structure, to empower ANNs.

\begin{table}[thb]
\centering
\begin{tabularx}{\textwidth}{@{} X X X @{} }
	\toprule
	 & \textbf{Artificial neurons in ANNs} & \textbf{Biological neurons} \\
	\midrule
	\textbf{Morphology} & Nodes as point neurons, each node can be a multilayered perceptron, symmetric tree & Complex dendritic tree with several branch points, not symmetric \\ \\
	\textbf{Sparsification} & Dropout layers & Inputs are distributed across the tree \\ \\
	\textbf{Integration} & Activation functions used in each node (e.g., sigmoid, tanh, ReLU) & Dendritic spikes (Na\textsuperscript{+}, Ca\textsuperscript{2+}, NMDA, dCaAPs, back-propagating action potentials) or passive (sublinear) integration \\ \\
	\textbf{Learning} & Backpropagation-of-error algorithm (most effective) & Cooperative synaptic plasticity (local), plasticity of excitability (dendritic/somatic), bursting \\ \\
	\textbf{Homeostasis} & Regularization of weights / Normalization of layers & The total synaptic strength of all synapses (in a branch, neuron, or network) scales up/down \\ \\
	\textbf{Rewiring} & Not typically used, would amount to dynamically changing the graph & Synapses turn over dynamically, and only those that fire with neighbors are stabilized \\ \\
	\bottomrule
\end{tabularx}
\label{tab:table1}
\end{table}

\section{Conclusions}
\label{sec:concl}

This opinion article highlights specific features of biological dendritic trees whose adoption is likely to endow important advantages to ANNs (summarized in Table \ref{tab:table1}). These features include a) the compartmentalizing and filtering effects of dendritic morphology, which effectively sparsify the connectivity of biological networks and form semi-independent processing units within individual neurons. b) The integrative properties of dendrites, culminating in the generation of dendritic spikes. These spikes transform synaptic inputs in ways similar to those of nonlinear activation functions in artificial nodes while also allowing dendrites to solve difficult spatio-temporal problems, such as logical operations (e.g., AND, OR, XOR, NAND), implement coincidence and sequence detection, associate input streams, etc. c) The dampening effects of morphology and the amplifying effects of dendritic spikes are bound together by a variety of plasticity rules that fine-tune input-output transformations at all levels of processing: from the strength of individual synapses to the grouping of correlated inputs, the coupling strength of each dendrite to its soma, the modulation of excitability and the homeostatic scaling of connections. We believe that these features boost learning and problem-solving in biological brains, not in isolation but through their intricate interplay. They do so by augmenting flexibility so that biological networks can quickly tune into new problems without forgetting the ones learned. This is realized via local updates that are not driven by prediction errors (which are task-specific) but rather by the regularities in the input streams (which can be generic across tasks). Updates occur within specific, sparsely connected compartments, thus allowing different problems to tab onto common features as detected at the bottom layers. This hypothesis is supported by the studies discussed here. Yet, it remains unknown whether dendrites can empower learning in ANNs in a manner that goes well beyond the traditional architectures, enabling them to solve more challenging problems.

What would certainly comprise an important advantage is the physical implementation of dendritic ANNs on neuromorphic chips. Currently, the power consumption of traditional DL architectures is much higher than that of a human brain, while their robustness and flexibility in different tasks remain limited \cite{Tang2019}. Neuromorphic computing, on the other hand, can perform complex calculations faster, more power-efficiently, and on a smaller area than traditional von Neumann architectures \cite{Markovic2020}. Despite the current advances in the hardware domain, e.g., chips stack in a 3D structure \cite{Lin2020}, the physical dimensions of chips and the need for long wires to emulate the all-to-all connectivity pose severe limitations in implementing state-of-the-art DL architectures on hardware. Dendritic ANNs offer a solution to this problem, as they massively reduce the need for long-cables and all-to-all connectivity \cite{Hussain2015,Roy2014}. Recent attempts to integrate dendritic functions on hardware \cite{Li2020} have resulted in a single chip containing the three critical components of a biological neuron; soma, dendrite, and synapses, which can efficiently process spatio-temporal information. These first steps are highly promising towards making neuromorphic computing a practical reality. We hope that the seeds planted here will inspire generations of ML and hardware researchers, helping them achieve major savings in resource utilization and perhaps even human-level performance in tasks that remain unsolvable by current systems.

\section{Author Contributions}
\label{sec:acontrib}
S.C. and P.P. conceived and wrote the paper.

\section{Conflict of interest statement}
\label{sec:conflict}
The authors declare no conflict of interest.

\section{Acknowledgments}
\label{sec:acknow}
We want to thank Blake Richards, Konrad Kording, and Richard Naud for their thoughtful comments and efforts towards improving our manuscript. This work was supported by the NEUREKA grant (H2020-FETOPEN-2018-2019-2020-01, FET-Open Challenging Current Thinking, GA-863245) and the Einstein Foundation Berlin (visiting fellowship to P.P.).

\bibliography{references}

\begin{thebibliography}{100}

\bibitem{LeCun2015}
Y.~LeCun, Y.~Bengio, and G.~Hinton, ``Deep learning,'' {\em Nature}, vol.~521,
  no.~7553, pp.~436--444, 2015.

\bibitem{Nassif2019}
A.~B. Nassif, I.~Shahin, I.~Attili, M.~Azzeh, and K.~Shaalan, ``Speech
  recognition using deep neural networks: A systematic review,'' {\em IEEE
  access}, vol.~7, pp.~19143--19165, 2019.

\bibitem{Masi2018}
I.~Masi, Y.~Wu, T.~Hassner, and P.~Natarajan, ``Deep face recognition: A
  survey,'' in {\em 2018 31st SIBGRAPI conference on graphics, patterns and
  images (SIBGRAPI)}, pp.~471--478, IEEE, 2018.

\bibitem{Justesen2020}
N.~Justesen, P.~Bontrager, J.~Togelius, and S.~Risi, ``Deep learning for video
  game playing,'' {\em IEEE Transactions on Games}, vol.~12, no.~1, pp.~1--20,
  2019.

\bibitem{Huang2020}
Y.~Huang and Y.~Chen, ``Survey of state-of-art autonomous driving technologies
  with deep learning,'' in {\em 2020 IEEE 20th International Conference on
  Software Quality, Reliability and Security Companion (QRS-C)}, pp.~221--228,
  IEEE, 2020.

\bibitem{Stuart2016}
G.~Stuart, N.~Spruston, and M.~H{\"a}usser, {\em Dendrites}.
\newblock Oxford University Press, 2016.

\bibitem{Spruston1994}
N.~Spruston, D.~B. Jaffe, and D.~Johnston, ``Dendritic attenuation of synaptic
  potentials and currents: the role of passive membrane properties,'' {\em
  Trends in neurosciences}, vol.~17, no.~4, pp.~161--166, 1994.

\bibitem{Major2013}
G.~Major, M.~E. Larkum, and J.~Schiller, ``Active properties of neocortical
  pyramidal neuron dendrites,'' {\em Annual review of neuroscience}, vol.~36,
  pp.~1--24, 2013.

\bibitem{Ariav2003}
G.~Ariav, A.~Polsky, and J.~Schiller, ``Submillisecond precision of the
  input-output transformation function mediated by fast sodium dendritic spikes
  in basal dendrites of ca1 pyramidal neurons,'' {\em Journal of Neuroscience},
  vol.~23, no.~21, pp.~7750--7758, 2003.

\bibitem{Branco2011}
T.~Branco and M.~H{\"a}usser, ``Synaptic integration gradients in single
  cortical pyramidal cell dendrites,'' {\em Neuron}, vol.~69, no.~5,
  pp.~885--892, 2011.

\bibitem{Polsky2004}
A.~Polsky, B.~W. Mel, and J.~Schiller, ``Computational subunits in thin
  dendrites of pyramidal cells,'' {\em Nature neuroscience}, vol.~7, no.~6,
  pp.~621--627, 2004.

\bibitem{Softky1994}
W.~Softky, ``Sub-millisecond coincidence detection in active dendritic trees,''
  {\em Neuroscience}, vol.~58, no.~1, pp.~13--41, 1994.

\bibitem{Spruston2008}
N.~Spruston, ``Pyramidal neurons: dendritic structure and synaptic
  integration,'' {\em Nature Reviews Neuroscience}, vol.~9, no.~3,
  pp.~206--221, 2008.

\bibitem{Ujfalussy2009}
B.~Ujfalussy, T.~Kiss, and P.~{\'E}rdi, ``Parallel computational subunits in
  dentate granule cells generate multiple place fields,'' {\em PLoS Comput
  Biol}, vol.~5, no.~9, p.~e1000500, 2009.

\bibitem{Ujfalussy2018}
B.~B. Ujfalussy, J.~K. Makara, M.~Lengyel, and T.~Branco, ``Global and
  multiplexed dendritic computations under in vivo-like conditions,'' {\em
  Neuron}, vol.~100, no.~3, pp.~579--592, 2018.

\bibitem{Bono2017}
J.~Bono, K.~A. Wilmes, and C.~Clopath, ``Modelling plasticity in dendrites:
  from single cells to networks,'' {\em Current opinion in neurobiology},
  vol.~46, pp.~136--141, 2017.

\bibitem{Kastellakis2019}
G.~Kastellakis and P.~Poirazi, ``Synaptic clustering and memory formation,''
  {\em Frontiers in molecular neuroscience}, vol.~12, p.~300, 2019.

\bibitem{Gordon2006}
U.~Gordon, A.~Polsky, and J.~Schiller, ``Plasticity compartments in basal
  dendrites of neocortical pyramidal neurons,'' {\em Journal of Neuroscience},
  vol.~26, no.~49, pp.~12717--12726, 2006.

\bibitem{Magee2020}
J.~C. Magee and C.~Grienberger, ``Synaptic plasticity forms and functions,''
  {\em Annual review of neuroscience}, vol.~43, pp.~95--117, 2020.

\bibitem{Mel2017}
B.~W. Mel, J.~Schiller, and P.~Poirazi, ``Synaptic plasticity in dendrites:
  complications and coping strategies,'' {\em Current opinion in neurobiology},
  vol.~43, pp.~177--186, 2017.

\bibitem{Sajikumar2011}
S.~Sajikumar and M.~Korte, ``Different compartments of apical ca1 dendrites
  have different plasticity thresholds for expressing synaptic tagging and
  capture,'' {\em Learning \& memory}, vol.~18, no.~5, pp.~327--331, 2011.

\bibitem{Weber2016}
J.~P. Weber, B.~K. Andr{\'a}sfalvy, M.~Polito, {\'A}.~Mag{\'o}, B.~B.
  Ujfalussy, and J.~K. Makara, ``Location-dependent synaptic plasticity rules
  by dendritic spine cooperativity,'' {\em Nature communications}, vol.~7,
  no.~1, pp.~1--14, 2016.

\bibitem{Losonczy2008}
A.~Losonczy, J.~K. Makara, and J.~C. Magee, ``Compartmentalized dendritic
  plasticity and input feature storage in neurons,'' {\em Nature}, vol.~452,
  no.~7186, pp.~436--441, 2008.

\bibitem{Jadi2014}
M.~P. Jadi, B.~F. Behabadi, A.~Poleg-Polsky, J.~Schiller, and B.~W. Mel, ``An
  augmented two-layer model captures nonlinear analog spatial integration
  effects in pyramidal neuron dendrites,'' {\em Proceedings of the IEEE},
  vol.~102, no.~5, pp.~782--798, 2014.

\bibitem{London2005}
M.~London and M.~H{\"a}usser, ``Dendritic computation,'' {\em Annu. Rev.
  Neurosci.}, vol.~28, pp.~503--532, 2005.

\bibitem{Poirazi2003a}
P.~Poirazi, T.~Brannon, and B.~W. Mel, ``Pyramidal neuron as two-layer neural
  network,'' {\em Neuron}, vol.~37, no.~6, pp.~989--999, 2003.

\bibitem{Tzilivaki2019}
A.~Tzilivaki, G.~Kastellakis, and P.~Poirazi, ``Challenging the point neuron
  dogma: Fs basket cells as 2-stage nonlinear integrators,'' {\em Nature
  communications}, vol.~10, no.~1, pp.~1--14, 2019.

\bibitem{Branco2010}
T.~Branco and M.~H{\"a}usser, ``The single dendritic branch as a fundamental
  functional unit in the nervous system,'' {\em Current opinion in
  neurobiology}, vol.~20, no.~4, pp.~494--502, 2010.

\bibitem{Guerguiev2017}
J.~Guerguiev, T.~P. Lillicrap, and B.~A. Richards, ``Towards deep learning with
  segregated dendrites,'' {\em eLife}, vol.~6, p.~e22901, 2017.

\bibitem{Kreiman2020}
G.~Kreiman and T.~Serre, ``Beyond the feedforward sweep: feedback computations
  in the visual cortex,'' {\em Annals of the New York Academy of Sciences},
  vol.~1464, no.~1, p.~222, 2020.

\bibitem{Payeur2021}
A.~Payeur, J.~Guerguiev, F.~Zenke, B.~A. Richards, and R.~Naud,
  ``Burst-dependent synaptic plasticity can coordinate learning in hierarchical
  circuits,'' {\em Nature Neuroscience}, pp.~1--10, 2021.

\bibitem{Siegel2000}
M.~Siegel, K.~P. K{\"o}rding, and P.~K{\"o}nig, ``Integrating top-down and
  bottom-up sensory processing by somato-dendritic interactions,'' {\em Journal
  of computational neuroscience}, vol.~8, no.~2, pp.~161--173, 2000.

\bibitem{Kirkpatrick2017}
J.~Kirkpatrick, R.~Pascanu, N.~Rabinowitz, J.~Veness, G.~Desjardins, A.~A.
  Rusu, K.~Milan, J.~Quan, T.~Ramalho, A.~Grabska-Barwinska, {\em et~al.},
  ``Overcoming catastrophic forgetting in neural networks,'' {\em Proceedings
  of the national academy of sciences}, vol.~114, no.~13, pp.~3521--3526, 2017.

\bibitem{Kording2001}
K.~P. K{\"o}rding and P.~K{\"o}nig, ``Supervised and unsupervised learning with
  two sites of synaptic integration,'' {\em Journal of computational
  neuroscience}, vol.~11, no.~3, pp.~207--215, 2001.

\bibitem{Zador1995}
A.~M. Zador, H.~Agmon-Snir, and I.~Segev, ``The morphoelectrotonic transform: a
  graphical approach to dendritic function,'' {\em Journal of Neuroscience},
  vol.~15, no.~3, pp.~1669--1682, 1995.

\bibitem{Stuart1998}
G.~Stuart and N.~Spruston, ``Determinants of voltage attenuation in neocortical
  pyramidal neuron dendrites,'' {\em Journal of Neuroscience}, vol.~18, no.~10,
  pp.~3501--3510, 1998.

\bibitem{Baruah2019}
S.~M.~B. Baruah, D.~Nandi, and S.~Roy, ``Modelling signal transmission in
  passive dendritic fibre using discretized cable equation,'' in {\em 2019 2nd
  International Conference on Innovations in Electronics, Signal Processing and
  Communication (IESC)}, pp.~138--141, IEEE, 2019.

\bibitem{Ferrante2013}
M.~Ferrante, M.~Migliore, and G.~A. Ascoli, ``Functional impact of dendritic
  branch-point morphology,'' {\em Journal of Neuroscience}, vol.~33, no.~5,
  pp.~2156--2165, 2013.

\bibitem{Rall1964}
W.~Rall, ``Theoretical significance of dendritic trees for neuronal
  input-output relations,'' {\em Neural theory and modeling}, pp.~73--97, 1964.

\bibitem{Caze2013}
R.~D. Caz{\'e}, M.~Humphries, and B.~Gutkin, ``Passive dendrites enable single
  neurons to compute linearly non-separable functions,'' {\em PLoS Comput
  Biol}, vol.~9, no.~2, p.~e1002867, 2013.

\bibitem{Brunel2004}
N.~Brunel, V.~Hakim, P.~Isope, J.-P. Nadal, and B.~Barbour, ``Optimal
  information storage and the distribution of synaptic weights: perceptron
  versus purkinje cell,'' {\em Neuron}, vol.~43, no.~5, pp.~745--757, 2004.

\bibitem{Chavlis2017}
S.~Chavlis, P.~C. Petrantonakis, and P.~Poirazi, ``Dendrites of dentate gyrus
  granule cells contribute to pattern separation by controlling sparsity,''
  {\em Hippocampus}, vol.~27, no.~1, pp.~89--110, 2017.

\bibitem{Schweighofer2001}
N.~Schweighofer, K.~Doya, and F.~Lay, ``Unsupervised learning of granule cell
  sparse codes enhances cerebellar adaptive control,'' {\em Neuroscience},
  vol.~103, no.~1, pp.~35--50, 2001.

\bibitem{Hawkins2016}
J.~Hawkins and S.~Ahmad, ``Why neurons have thousands of synapses, a theory of
  sequence memory in neocortex,'' {\em Frontiers in neural circuits}, vol.~10,
  p.~23, 2016.

\bibitem{McCulloch1943}
W.~S. McCulloch and W.~Pitts, ``A logical calculus of the ideas immanent in
  nervous activity,'' {\em The bulletin of mathematical biophysics}, vol.~5,
  no.~4, pp.~115--133, 1943.

\bibitem{Srivastava2014}
N.~Srivastava, G.~Hinton, A.~Krizhevsky, I.~Sutskever, and R.~Salakhutdinov,
  ``Dropout: A simple way to prevent neural networks from overfitting,'' {\em
  Journal of Machine Learning Research}, vol.~15, no.~56, pp.~1929--1958, 2014.

\bibitem{Poirazi2004}
P.~Poirazi, C.~Neocleous, C.~S. Pattichis, and C.~N. Schizas, ``Classification
  capacity of a modular neural network implementing neurally inspired
  architecture and training rules,'' {\em IEEE Transactions on Neural
  Networks}, vol.~15, no.~3, pp.~597--612, 2004.

\bibitem{Shukla2010}
A.~Shukla, R.~Tiwari, and R.~Kala, ``Modular neural networks,'' in {\em Towards
  Hybrid and Adaptive Computing. Studies in Computational Intelligence},
  vol.~307, pp.~307--335, Berlin, Heidelberg: Springer-Verlag, 2010.

\bibitem{BaoyuanLiu2015}
B.~Liu, M.~Wang, H.~Foroosh, M.~Tappen, and M.~Pensky, ``Sparse convolutional
  neural networks,'' in {\em Proceedings of the IEEE conference on computer
  vision and pattern recognition}, pp.~806--814, 2015.

\bibitem{Losonczy2006}
A.~Losonczy and J.~C. Magee, ``Integrative properties of radial oblique
  dendrites in hippocampal ca1 pyramidal neurons,'' {\em Neuron}, vol.~50,
  no.~2, pp.~291--307, 2006.

\bibitem{Poirazi2003}
P.~Poirazi, T.~Brannon, and B.~W. Mel, ``Arithmetic of subthreshold synaptic
  summation in a model ca1 pyramidal cell,'' {\em Neuron}, vol.~37, no.~6,
  pp.~977--987, 2003.

\bibitem{Poirazi2001}
P.~Poirazi and B.~W. Mel, ``Impact of active dendrites and structural
  plasticity on the memory capacity of neural tissue,'' {\em Neuron}, vol.~29,
  no.~3, pp.~779--796, 2001.

\bibitem{Golding1998}
N.~L. Golding and N.~Spruston, ``Dendritic sodium spikes are variable triggers
  of axonal action potentials in hippocampal ca1 pyramidal neurons,'' {\em
  Neuron}, vol.~21, no.~5, pp.~1189--1200, 1998.

\bibitem{Milojkovic2005}
B.~Milojkovic, J.~Wuskell, L.~Loew, and S.~Antic, ``Initiation of sodium
  spikelets in basal dendrites of neocortical pyramidal neurons,'' {\em The
  Journal of membrane biology}, vol.~208, no.~2, pp.~155--169, 2005.

\bibitem{Larkum2009}
M.~E. Larkum, T.~Nevian, M.~Sandler, A.~Polsky, and J.~Schiller, ``Synaptic
  integration in tuft dendrites of layer 5 pyramidal neurons: a new unifying
  principle,'' {\em Science}, vol.~325, no.~5941, pp.~756--760, 2009.

\bibitem{Schiller1997}
J.~Schiller, Y.~Schiller, G.~Stuart, and B.~Sakmann, ``Calcium action
  potentials restricted to distal apical dendrites of rat neocortical pyramidal
  neurons,'' {\em The Journal of physiology}, vol.~505, no.~3, pp.~605--616,
  1997.

\bibitem{Cichon2015}
J.~Cichon and W.-B. Gan, ``Branch-specific dendritic ca 2+ spikes cause
  persistent synaptic plasticity,'' {\em Nature}, vol.~520, no.~7546,
  pp.~180--185, 2015.

\bibitem{Brunner2016}
J.~Brunner and J.~Szabadics, ``Analogue modulation of back-propagating action
  potentials enables dendritic hybrid signalling,'' {\em Nature
  communications}, vol.~7, no.~1, pp.~1--13, 2016.

\bibitem{Silver2010}
R.~A. Silver, ``Neuronal arithmetic,'' {\em Nature Reviews Neuroscience},
  vol.~11, no.~7, pp.~474--489, 2010.

\bibitem{Larkum1999}
M.~E. Larkum, J.~J. Zhu, and B.~Sakmann, ``A new cellular mechanism for
  coupling inputs arriving at different cortical layers,'' {\em Nature},
  vol.~398, no.~6725, pp.~338--341, 1999.

\bibitem{Beniaguev2020}
D.~Beniaguev, I.~Segev, and M.~London, ``Single cortical neurons as deep
  artificial neural networks,'' {\em bioRxiv}, p.~613141, 2020.

\bibitem{Hausser2003}
M.~H{\"a}usser and B.~Mel, ``Dendrites: bug or feature?,'' {\em Current opinion
  in neurobiology}, vol.~13, no.~3, pp.~372--383, 2003.

\bibitem{Chiovini2014}
B.~Chiovini, G.~F. Turi, G.~Katona, A.~Kasz{\'a}s, D.~P{\'a}lfi, P.~Ma{\'a}k,
  G.~Szalay, M.~F. Szab{\'o}, G.~Szab{\'o}, Z.~Szadai, {\em et~al.},
  ``Dendritic spikes induce ripples in parvalbumin interneurons during
  hippocampal sharp waves,'' {\em Neuron}, vol.~82, no.~4, pp.~908--924, 2014.

\bibitem{Katona2011}
G.~Katona, A.~Kasz{\'a}s, G.~F. Turi, N.~H{\'a}jos, G.~Tam{\'a}s, E.~S. Vizi,
  and B.~R{\'o}zsa, ``Roller coaster scanning reveals spontaneous triggering of
  dendritic spikes in ca1 interneurons,'' {\em Proceedings of the National
  Academy of Sciences}, vol.~108, no.~5, pp.~2148--2153, 2011.

\bibitem{Tran-Van-Minh2016}
A.~Tran-Van-Minh, T.~Abrahamsson, L.~Cathala, and D.~A. DiGregorio,
  ``Differential dendritic integration of synaptic potentials and calcium in
  cerebellar interneurons,'' {\em Neuron}, vol.~91, no.~4, pp.~837--850, 2016.

\bibitem{Gidon2020}
A.~Gidon, T.~A. Zolnik, P.~Fidzinski, F.~Bolduan, A.~Papoutsi, P.~Poirazi,
  M.~Holtkamp, I.~Vida, and M.~E. Larkum, ``Dendritic action potentials and
  computation in human layer 2/3 cortical neurons,'' {\em Science}, vol.~367,
  no.~6473, pp.~83--87, 2020.

\bibitem{Hayou2019}
S.~Hayou, A.~Doucet, and J.~Rousseau, ``On the impact of the activation
  function on deep neural networks training,'' in {\em International Conference
  on Machine Learning}, pp.~2672--2680, PMLR, 2019.

\bibitem{Wu2018}
X.~Wu, X.~Liu, W.~Li, and Q.~Wu, ``Improved expressivity through dendritic
  neural networks,'' in {\em Proceedings of the 32nd International Conference
  on Neural Information Processing Systems}, pp.~8068--8079, 2018.

\bibitem{Jones2020}
I.~S. Jones and K.~P. Kording, ``Can single neurons solve mnist? the
  computational power of biological dendritic trees,'' {\em arXiv preprint
  arXiv:2009.01269}, 2020.

\bibitem{Camp2020}
B.~Camp, J.~K. Mandivarapu, and R.~Estrada, ``Continual learning with deep
  artificial neurons,'' {\em arXiv preprint arXiv:2011.07035}, 2020.

\bibitem{Georgescu2020}
M.-I. Georgescu, R.~T. Ionescu, N.-C. Ristea, and N.~Sebe, ``Non-linear neurons
  with human-like apical dendrite activations,'' {\em arXiv preprint
  arXiv:2003.03229}, 2020.

\bibitem{Bittner2017}
K.~C. Bittner, A.~D. Milstein, C.~Grienberger, S.~Romani, and J.~C. Magee,
  ``Behavioral time scale synaptic plasticity underlies ca1 place fields,''
  {\em Science}, vol.~357, no.~6355, pp.~1033--1036, 2017.

\bibitem{Kampa2007}
B.~M. Kampa, J.~J. Letzkus, and G.~J. Stuart, ``Dendritic mechanisms
  controlling spike-timing-dependent synaptic plasticity,'' {\em Trends in
  neurosciences}, vol.~30, no.~9, pp.~456--463, 2007.

\bibitem{El-Boustani2018}
S.~El-Boustani, J.~P. Ip, V.~Breton-Provencher, G.~W. Knott, H.~Okuno, H.~Bito,
  and M.~Sur, ``Locally coordinated synaptic plasticity of visual cortex
  neurons in vivo,'' {\em Science}, vol.~360, no.~6395, pp.~1349--1354, 2018.

\bibitem{Kastellakis2016}
G.~Kastellakis, A.~J. Silva, and P.~Poirazi, ``Linking memories across time via
  neuronal and dendritic overlaps in model neurons with active dendrites,''
  {\em Cell reports}, vol.~17, no.~6, pp.~1491--1504, 2016.

\bibitem{Lee2016}
K.~F. Lee, C.~Soares, J.-P. Thivierge, and J.-C. B{\'e}{\"\i}que, ``Correlated
  synaptic inputs drive dendritic calcium amplification and cooperative
  plasticity during clustered synapse development,'' {\em Neuron}, vol.~89,
  no.~4, pp.~784--799, 2016.

\bibitem{Chklovskii2004}
D.~B. Chklovskii, B.~Mel, and K.~Svoboda, ``Cortical rewiring and information
  storage,'' {\em Nature}, vol.~431, no.~7010, pp.~782--788, 2004.

\bibitem{DeBello2008}
W.~M. DeBello, ``Micro-rewiring as a substrate for learning,'' {\em Trends in
  neurosciences}, vol.~31, no.~11, pp.~577--584, 2008.

\bibitem{DeBello2017}
W.~DeBello and K.~Zito, ``Within a spine’s reach,'' in {\em The Rewiring
  Brain}, pp.~295--317, Elsevier, 2017.

\bibitem{Legenstein2011}
R.~Legenstein and W.~Maass, ``Branch-specific plasticity enables
  self-organization of nonlinear computation in single neurons,'' {\em Journal
  of Neuroscience}, vol.~31, no.~30, pp.~10787--10802, 2011.

\bibitem{Limbacher2020}
T.~Limbacher and R.~Legenstein, ``Emergence of stable synaptic clusters on
  dendrites through synaptic rewiring,'' {\em Frontiers in computational
  neuroscience}, vol.~14, p.~57, 2020.

\bibitem{Turrigiano2008}
G.~G. Turrigiano, ``The self-tuning neuron: synaptic scaling of excitatory
  synapses,'' {\em Cell}, vol.~135, no.~3, pp.~422--435, 2008.

\bibitem{Rabinowitch2006}
I.~Rabinowitch and I.~Segev, ``The interplay between homeostatic synaptic
  plasticity and functional dendritic compartments,'' {\em Journal of
  neurophysiology}, vol.~96, no.~1, pp.~276--283, 2006.

\bibitem{Disterhoft2006}
J.~F. Disterhoft and M.~M. Oh, ``Learning, aging and intrinsic neuronal
  plasticity,'' {\em Trends in neurosciences}, vol.~29, no.~10, pp.~587--599,
  2006.

\bibitem{Ohtsuki2012}
G.~Ohtsuki, C.~Piochon, J.~P. Adelman, and C.~Hansel, ``Sk2 channel modulation
  contributes to compartment-specific dendritic plasticity in cerebellar
  purkinje cells,'' {\em Neuron}, vol.~75, no.~1, pp.~108--120, 2012.

\bibitem{Makara2009}
J.~K. Makara, A.~Losonczy, Q.~Wen, and J.~C. Magee, ``Experience-dependent
  compartmentalized dendritic plasticity in rat hippocampal ca1 pyramidal
  neurons,'' {\em Nature neuroscience}, vol.~12, no.~12, p.~1485, 2009.

\bibitem{Cai2016}
D.~J. Cai, D.~Aharoni, T.~Shuman, J.~Shobe, J.~Biane, W.~Song, B.~Wei,
  M.~Veshkini, M.~La-Vu, J.~Lou, {\em et~al.}, ``A shared neural ensemble links
  distinct contextual memories encoded close in time,'' {\em Nature}, vol.~534,
  no.~7605, pp.~115--118, 2016.

\bibitem{Kirchner2019}
J.~H. Kirchner and J.~Gjorgjieva, ``A unifying framework for synaptic
  organization on cortical dendrites,'' {\em BioRxiv}, p.~771907, 2019.

\bibitem{Bellec2018}
G.~Bellec, D.~Kappel, W.~Maass, and R.~Legenstein, ``Deep rewiring: Training
  very sparse deep networks,'' {\em arXiv preprint arXiv:1711.05136}, 2017.

\bibitem{Doron2020}
G.~Doron, J.~N. Shin, N.~Takahashi, M.~Dr{\"u}ke, C.~Bocklisch, S.~Skenderi,
  L.~de~Mont, M.~Toumazou, J.~Ledderose, M.~Brecht, {\em et~al.}, ``Perirhinal
  input to neocortical layer 1 controls learning,'' {\em Science}, vol.~370,
  no.~6523, 2020.

\bibitem{Illing2019}
B.~Illing, W.~Gerstner, and J.~Brea, ``Biologically plausible deep
  learning—but how far can we go with shallow networks?,'' {\em Neural
  Networks}, vol.~118, pp.~90--101, 2019.

\bibitem{Perez-Sanchez2018}
B.~P{\'e}rez-S{\'a}nchez, O.~Fontenla-Romero, and B.~Guijarro-Berdi{\~n}as, ``A
  review of adaptive online learning for artificial neural networks,'' {\em
  Artificial Intelligence Review}, vol.~49, no.~2, pp.~281--299, 2018.

\bibitem{Lillicrap2016}
T.~P. Lillicrap, D.~Cownden, D.~B. Tweed, and C.~J. Akerman, ``Random synaptic
  feedback weights support error backpropagation for deep learning,'' {\em
  Nature communications}, vol.~7, no.~1, pp.~1--10, 2016.

\bibitem{Lillicrap2020}
T.~P. Lillicrap, A.~Santoro, L.~Marris, C.~J. Akerman, and G.~Hinton,
  ``Backpropagation and the brain,'' {\em Nature Reviews Neuroscience},
  vol.~21, no.~6, pp.~335--346, 2020.

\bibitem{Richards2019}
B.~A. Richards, T.~P. Lillicrap, P.~Beaudoin, Y.~Bengio, R.~Bogacz,
  A.~Christensen, C.~Clopath, R.~P. Costa, A.~de~Berker, S.~Ganguli, {\em
  et~al.}, ``A deep learning framework for neuroscience,'' {\em Nature
  neuroscience}, vol.~22, no.~11, pp.~1761--1770, 2019.

\bibitem{Sacramento2018}
J.~Sacramento, R.~P. Costa, Y.~Bengio, and W.~Senn, ``Dendritic cortical
  microcircuits approximate the backpropagation algorithm,'' {\em arXiv
  preprint arXiv:1810.11393}, 2018.

\bibitem{Mostafa2018}
H.~Mostafa, V.~Ramesh, and G.~Cauwenberghs, ``Deep supervised learning using
  local errors,'' {\em Frontiers in neuroscience}, vol.~12, p.~608, 2018.

\bibitem{Nokland2019}
A.~N{\o}kland and L.~H. Eidnes, ``Training neural networks with local error
  signals,'' pp.~4839--4850, 2019.

\bibitem{Lee2015}
D.-H. Lee, S.~Zhang, A.~Fischer, and Y.~Bengio, ``Difference target
  propagation,'' in {\em Joint european conference on machine learning and
  knowledge discovery in databases}, pp.~498--515, Springer, 2015.

\bibitem{Choromanska2019}
A.~Choromanska, B.~Cowen, S.~Kumaravel, R.~Luss, M.~Rigotti, I.~Rish,
  P.~Diachille, V.~Gurev, B.~Kingsbury, R.~Tejwani, {\em et~al.}, ``Beyond
  backprop: Online alternating minimization with auxiliary variables,'' in {\em
  International Conference on Machine Learning}, pp.~1193--1202, PMLR, 2019.

\bibitem{Lipshutz2020}
D.~Lipshutz, Y.~Bahroun, S.~Golkar, A.~M. Sengupta, and D.~B. Chkovskii, ``A
  biologically plausible neural network for multi-channel canonical correlation
  analysis,'' {\em arXiv preprint arXiv:2010.00525}, 2020.

\bibitem{Illing2020}
B.~Illing, W.~Gerstner, and G.~Bellec, ``Towards truly local gradients with
  clapp: Contrastive, local and predictive plasticity,'' {\em arXiv preprint
  arXiv:2010.08262}, 2020.

\bibitem{Tang2019}
J.~Tang, F.~Yuan, X.~Shen, Z.~Wang, M.~Rao, Y.~He, Y.~Sun, X.~Li, W.~Zhang,
  Y.~Li, {\em et~al.}, ``Bridging biological and artificial neural networks
  with emerging neuromorphic devices: fundamentals, progress, and challenges,''
  {\em Advanced Materials}, vol.~31, no.~49, p.~1902761, 2019.

\bibitem{Markovic2020}
D.~Markovi{\'c}, A.~Mizrahi, D.~Querlioz, and J.~Grollier, ``Physics for
  neuromorphic computing,'' {\em Nature Reviews Physics}, vol.~2, no.~9,
  pp.~499--510, 2020.

\bibitem{Lin2020}
P.~Lin, C.~Li, Z.~Wang, Y.~Li, H.~Jiang, W.~Song, M.~Rao, Y.~Zhuo, N.~K.
  Upadhyay, M.~Barnell, {\em et~al.}, ``Three-dimensional memristor circuits as
  complex neural networks,'' {\em Nature Electronics}, vol.~3, no.~4,
  pp.~225--232, 2020.

\bibitem{Hussain2015}
S.~Hussain, S.-C. Liu, and A.~Basu, ``Hardware-amenable structural learning for
  spike-based pattern classification using a simple model of active
  dendrites,'' {\em Neural computation}, vol.~27, no.~4, pp.~845--897, 2015.

\bibitem{Roy2014}
S.~Roy, A.~Banerjee, and A.~Basu, ``Liquid state machine with dendritically
  enhanced readout for low-power, neuromorphic vlsi implementations,'' {\em
  IEEE transactions on biomedical circuits and systems}, vol.~8, no.~5,
  pp.~681--695, 2014.

\bibitem{Li2020}
X.~Li, B.~Yu, B.~Wang, L.~Bao, B.~Zhang, H.~Li, Z.~Yu, T.~Zhang, Y.~Yang,
  R.~Huang, {\em et~al.}, ``Multi-terminal ionic-gated low-power silicon
  nanowire synaptic transistors with dendritic functions for neuromorphic
  systems,'' {\em Nanoscale}, vol.~12, no.~30, pp.~16348--16358, 2020.

\end{thebibliography}
\bibliographystyle{ieeetr}

\end{document}